\documentclass[aps,twocolumn,aps,floatfix]{revtex4}
\usepackage{graphicx}
\usepackage{bm}
\usepackage{hyperref}
\def\hm2{\frac{\hbar^2}{2m}}
\def\vecr{ {\bf r}}
\def\veck{ {\bf k}}
\def\he4{$^4$He} 

\begin{document}
\title{Excitations in confined helium}
\author{V. Apaja$^{1,2}$ and E. Krotscheck$^1$}
\affiliation{$^1$Institut f\"ur Theoretische Physik, Johannes Kepler
Universit\"at, A 4040 Linz, Austria}
\affiliation{$^2$Department of
Physical Sciences, Theoretical Physics, University of Oulu, FIN-90014
Oulu, Finland}

\begin{abstract}
We design models for helium in matrices like aerogel, Vycor or Geltech
from a manifestly microscopic point of view. For that purpose, we
calculate the dynamic structure function of $^4$He on Si substrates
and between two Si walls as a function of energy, momentum transfer,
and the scattering angle. The angle--averaged results are in good
agreement with the neutron scattering data; the remaining differences
can be attributed to the simplified model used here for the complex
pore structure of the materials. A focus of the present work is the
detailed identification of coexisting layer modes and bulk--like
excitations, and, in the case of thick films, ripplon
excitations. Involving essentially two--dimensional motion of atoms,
the layer modes are sensitive to the scattering angle.

\end{abstract}
\maketitle

\section{introduction}

There is much current interest in understanding the properties of
superfluid $^4$He in confinement. In particular, collective
excitations of superfluid helium confined to silica aerogel have been
studied by neutron scattering since the early 90's
\cite{snow-sokol-90,kinder-goddens-millet-94}, and by now a wealth of
information about helium in aerogel, Vycor and Geltech has been
collected \cite{dimeo-etal-97,dimeo-sokol-anderson-stirling-adams-98,%
sokol-gibbs-stirling-azuah-adams-96,gibbs-sokol-stirling-azuah-adams-97,%
plantevin-fak-glyde-bossy-beamish-98,glyde-fak-plantevin-98,%
fak-plantevin-glyde-00,plantevin-etal-01,plantevin-etal-02}. Aerogel
is an open gel structure formed by silica strands (SiO$_2$).  Typical
pore sizes range from few \AA\ to few hundred \AA, without any
characteristic pore size. Vycor is a porous glass, where pores form
channels of about 70~\AA\ diameter. Geltech resembles aerogel, except
that the nominal pore size is 25~\AA\ \cite{plantevin-etal-02}. When
liquid helium is placed in such an environment, it will first be
physisorbed to the free surfaces of the matrices. Such an adsorption
occurs in layers, the first layer of $^4$He is expected to be solid;
on a more strongly binding substrate, such as graphite, one expects
two solid layers. In this work we are restricted to the liquid state,
and the layer of solid helium is considered as part of the substrate.

Recent very accurate instruments have, contrary to earlier findings,
revealed that the energies and lifetimes of phonon--roton excitations
for confined $^4$He are nearly equal to their bulk superfluid $^4$He
values. Specifically, the roton linewidth was found by Anderson {\em
et al.}~\cite{anderson-etal-99} to be less than the instrumental
resolution, 0.1$\mu$m. As expected, differences usually appear at
partial fillings. The appearance of ripplons is tied to the existence
of a free liquid surface; neutron scattering experiments show clearly
their presence in on adsorbed films \cite{LauterJLTP,LauterPRL} and in
aerogel containing few layers of helium
\cite{bauerle-ILL-report-96,godfrin-ILL-report-98}.

Since nothing spectacular could be seen in the bulk--like mode, the
focus shifted to nearly two--dimensional ``layer modes''. The
existence of collective excitations below the roton minimum has been
demonstrated first by Lauter and collaborators
\cite{lauter-frank-godfrin-leiderer-89,
godfrin-frank-lauter-leiderer-89, lauter-godfrin-frank-leiderer-91,
lauter-godfrin-frank-leiderer-92}; these modes were identified as
longitudinal phonons propagating in the first liquid layer close to
the substrate \cite{ExcLett,filmexpt,filmexc}.  The nature of these
modes has been found to be very resilient and quite independent on the
substrate strength \cite{alkalis}, only for very weakly attractive
substrates like Cs, the ``layer phonons'' acquire a transverse
component and become similar to ripplons. Layer modes in helium have
been observed in Vycor \cite{glyde-etal-00} and in both fully filled
and partially filled aerogel above wave vector
1.7~\AA$^{-1}$~\cite{fak-plantevin-glyde-mulders-00}. Therefore, layer
modes are the only excitations observed that are {\it
characteristic\/} for helium films or helium in aerogel or Vycor.

The effect of disorder on macroscopic properties of both $^3$He and
$^4$He has been investigated in detail, but there are only few
theoretical studies on excitations that apply to confined
helium. Locally, the density profile of the liquid can be assumed to
be that of a helium film \cite{filmstruc} or helium filling a space
between two planes \cite{apaja-krotscheck-01b}. Impurity induced
disorder has been studied using Path Integral Monte Carlo
\cite{boninsegni-glyde-98}. At the moment there are no
close--to--reality models of the confining structures combined with a
proper theory of superfluid helium. Obvious simplified model systems
are liquid helium filling random voids or helium between parallel
planes. The former is computationally very demanding. Experiments have
covered the full range of helium thicknesses, from a bare monolayer to
a filled sample, and the results show that layer modes are visible
against the background of bulk excitations if there are about four or
five layers of helium; in those cases we are basically dealing with a
thick He film on Si. However, the pores in aerogel and Geltech are not
filling uniformly, which has lead to the notion of connected
superfluidity and localized Bose-Einstein condensation
\cite{plantevin-etal-02}. Thus a well--defined phonon--roton spectrum
may persist above the critical temperature.

\section{Theory}
\label{sec:theory}

The method of choice for studying the strongly interacting quantum
liquid $^4$He is the Jastrow--Feenberg theory. The theory is
microscopic in the sense that it begins with the best available
representation of the many-body Hamiltonian,
\begin{equation}
        H = \sum_i\left[-{\hbar^2\over 2m}\nabla_i^2
        + U_{\rm sub}(\vecr_i)\right]
        + \sum_{i<j} V(|\vecr_i-\vecr_j|)\ ,
\label{eq:hamiltonian}
\end{equation}
where $V(|\vecr_i-\vecr_j|)$ is the He-He
interaction \cite{aziz-mccourt-wong-87}. Compared to the bulk liquid,
the essential new property that must be dealt with is the breaking of
translational invariance and anisotropy of the system. Such symmetry
breakings may be induced by the substrate potential $U_{\rm
sub}(\vecr_i)$, but they may also occur spontaneously like in the
formation of quantum liquid droplets \cite{HNCdrops,dropdyn}. The
ground--state theory has been throughly discussed in
Ref.~\onlinecite{filmstruc} and the corresponding excited--state
theory is covered in Ref.~\onlinecite{filmexc}, therefore we only need
to review the key points.

The ground--state wave function
\begin{eqnarray}
&&  \Psi_0(\vecr_1,\ldots,\vecr_N)
        = \exp {1\over2}\Biggr[\sum_i u_1(\vecr_i)
        + \sum_{i<j} u_2(\vecr_i,\vecr_j)\nonumber\\
&&\hspace{3cm}      + \sum_{i<j<k}u_3(\vecr_i,\vecr_j,\vecr_k)+\ldots\Biggr]\,
\label{eq:wavefunction}
\end{eqnarray}
is written in {\it Jastrow-Feenberg\/} form; the $n$-body {\it
correlation functions\/} $u_n(\vecr_1,\ldots,\vecr_n)$ are determined
by functional minimization of the energy. The theory yields the
ground--state energetics and structure of the liquid and provides the
raw material for investigating excitations.

To introduce excitations to the system one applies a small,
time--dependent perturbation that momentarily drives the quantum
liquid out of its ground state. This causes the $n$-body correlations
in the wave function presented in Eq.~(\ref{eq:wavefunction}) to
acquire time dependence, hence the excited state has the form
\begin{equation}
\left|\Psi(t)\right\rangle = {e^{-iE_0 t/\hbar} \,
        e^{{1\over 2}\delta U(t)}\left|\Psi_0\right\rangle\over
        \left[\left\langle\Psi_0\left|e^{\,\Re e\delta U(t)}
        \right|\Psi_0\right\rangle\right]^{1/2}}
\end{equation}
with the {\it excitation operator}
\begin{equation}
\delta U(t) = \sum_i\delta u_1({\bf
r}_i;t) + \sum_{i<j} \delta u_2(\vecr_i,\vecr_j;t)+\ldots\,.
\label{eq:deltaU}
\end{equation}
The time--dependent correlation functions $\delta
u_n(\vecr_1,\ldots,\vecr_n;t)$ are determined by an
action principle \cite{kramer-saraceno-81}
\begin{equation}
        \delta \int ^{t_1}_{t_0} dt
        \left\langle \Psi (t) \left\vert H-i\hbar {\partial \over
\partial t} + U_{\rm ext}(t)\right\vert \Psi (t) \right\rangle   = 0\,,
\label{eq:S}
\end{equation}
where $U_{\rm ext}(t)$ is the weak external potential driving the
excitations.

Linearizing the equations of motion for the wave function
$\left|\Psi(t)\right\rangle =\left|\Psi_0\right\rangle +
\delta\left|\Psi(t)\right\rangle$ and calculating the time--dependent
component of the transition density $\delta\rho(\vecr,t) =
\bigl\langle \Psi_0\bigr| \hat\rho({\bf
r})\bigl|\delta\Psi(t)\bigr\rangle + c.c.$ allows us to calculate the
density--density response function $\chi(\vecr,\vecr';\omega)$
defined via
\begin{equation}
\delta \rho(\vecr,\omega)
= \int d^3 r' \chi(\vecr,\vecr';\omega)
U_{\rm ext}(\vecr',\omega)\,.
\end{equation}
Once the response function is known, one can apply the
fluctuation--dissipation theorem to find the dynamic structure function,
\begin{equation}
S(\vecr,\vecr';\omega) = -\frac{1}{\pi}\Im m  {\cal X}(\vecr,\vecr';\omega)\ .
\label{eq:Srrw}
\end{equation}
To obtain the dynamic structure function measured by, {\it e.g.},
neutron scattering, one has to project $S(\vecr,\vecr',\omega)$ onto
plane waves:
\begin{equation}
S(\veck,\omega) = \int d\vecr d\vecr' e^{i\veck\cdot(\vecr-\vecr')}
 S(\vecr,\vecr';\omega)\ .
\label{eq:Skw}
\end{equation}
Up to this points the formulas are valid for any geometry. Since the
systems with slit or slab geometry under consideration here are
translationally invariant only in the $(x-y)$ plane, but not in the
$z$ direction, only the momentum transfer $\veck_\|$ parallel to the
film or slab is a good quantum number. In this case , we have
\begin{equation}
S(\veck,\omega)_{\rm film} = \int dz dz' e^{i\veck_\perp(z-z')}
 S(\veck_\|,z,z';\omega)\,.
\label{eq:Sqwfilm}
\end{equation}

The truncation of the sequence of fluctuating correlations $\delta
u_n$ in Eq.~(\ref{eq:deltaU}) defines the level of approximation in
which we treat the excitations. The excitation spectrum can be quite
well understood \cite{JacksonSkw,Chuckphonon,VesaMikkou2} by retaining
only the time--dependent one-- and two--body terms in the excitation
operator (\ref{eq:deltaU}). The two--body terms $\delta
u_2(\vecr_1,\vecr_2;t)$ describe the interaction of two excitations.
The simplest non--trivial implementation of the theory
leads to a density--density response function of the form \cite{filmexc}
\begin{eqnarray}
        &&\chi({\bf r},{\bf r}',\omega) = \\
	&&\sqrt{\rho({\bf r})}
        \sum_{st}\phi^{(s)}({\bf r})
        \left[G_{st}(\omega)+
                G_{st}(-\omega)\right]
        \phi^{(t)}({\bf r}')\sqrt{\rho({\bf r}')}\nonumber
\label{eq:CBFresponse}
\end{eqnarray}
where the $\phi^{(s)}({\bf r})$ are Feynman excitation functions, and
\begin{equation}
G_{st}(\omega) =
\left[\hbar[\omega - \omega_s+i\epsilon]\delta_{st}
+ \Sigma_{st}(\omega)\right]^{-1}
\label{eq:Gdef}
\end{equation}
the phonon propagator. The fluctuating pair correlations
give rise to the self energy \cite{filmexc},
\begin{equation}
\Sigma_{st}(\omega) =
\frac{1}{2}\sum_{mn}\frac{V_{mn}^{(s)}V_{mn}^{(t)}}
{\hbar(\omega_m + \omega_n - \omega)}\,.
\label{eq:selfen}
\end{equation}
Here, the summation is over the Feynman states $m, n$; they form a partly
discrete, partly continuous set due to the inhomogeneity of the
liquid.  The expression for the three--phonon coupling amplitudes
$V_{mn}^{(s)}$ can be found in Ref.~\onlinecite{filmexc}. This self
energy renormalizes the Feynman ``phonon'' energies $\omega_n$, and
adds a finite lifetime to states that can decay to two lower--energy
modes. The resulting density--density response function has the
structure of a Brillouin-Wigner (BW) perturbation formula. The
approximate form of the self energy given in Eq.~(\ref{eq:selfen}) is
also closely related to the one obtained using the theory of
correlated basis functions (CBF)~\cite{JacksonSkw,Chuckphonon}.  As a
final refinement to the theory, we scale the Feynman energies
$\omega_n$ appearing in the energy denominator of the self energy given
in Eq.~(\ref{eq:selfen}) such that the roton minimum of the spectrum
used in the energy denominator of Eq. (\ref{eq:selfen}) agrees roughly
with the roton minimum predicted by the calculated $S({\bf k},
\omega)$. This is just a computationally simple way of adding the self
energy correction to the excitation energies in the self energy
itself. We shall use this scaled CBF-BW approximation for the
numerical parts of this paper.

A direct characterization of each mode is obtained by computing the
{\it transition density\/}
\begin{equation}
\delta\rho(\vecr;t) =
\left\langle\Psi_0\right|\hat\rho({\bf r})\left|\delta \Psi(t)\right\rangle
+ \hbox{c.c.}
\end{equation}
and the {\it transition current}
\begin{equation}
\delta{\bf j}(\vecr;t) =
\left\langle\Psi_0\right|\hat{\bf j}({\bf r})\left|\delta \Psi(t)\right\rangle
+ \hbox{c.c.}
\end{equation}
where $\hat\rho({\bf r})$ and $\hat{\bf j}({\bf r})$ are the familiar
one--body density and current operators. The transition density shows
the density change (arbitrary amplitude) and the transition current
the flow pattern of atoms in the mode.

\section{results}

We have modeled the confined quantum liquid by \he4 between two planar
substrates.  The distance between the substrate planes in our
calculations is 40~\AA\, which is not too far from the diameter of
aerogel strands and an intermediate value between the pore or channel
diameters in Vycor ($d\sim 70$\AA) and Geltech ($d\sim 25$\AA). We
assume translational invariance in a plane parallel to the surface.
This is, of course, not exactly true in the above mentioned
materials. However, our assumption should only change the details of
the excitations spectrum at long wave lengths that are comparable to
the pore size, whereas rotons arise from short range correlations.

\begin{figure}
\includegraphics[width=0.48\textwidth]{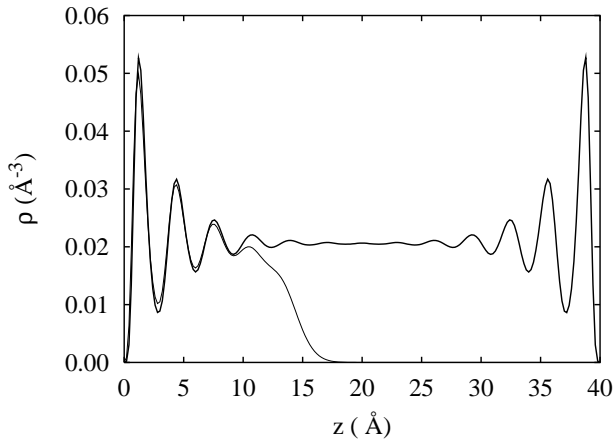}
\caption{The density profiles of the film and the filled gap used in
this work. The coverages of the film is 0.3~\AA$^{-2}$, and the filled
gap has 0.85 atoms per \AA${^2}$.}
\label{fig:rhos}
\end{figure}

We have computed the dynamic structure function as a function of the
scattering angle for a thick film and a filled gap. The density
profiles for these two situations are shown in
Fig.~\ref{fig:rhos}. The interaction of $^4$He particles with the
walls is described by the usual 3-9 potential obtained from averaging
Lennard-Jones potentials over a half space. This is valid if the walls
are smooth and, hence, $U_{\rm sub}(\vecr)$ is a function of one
coordinate only. Our approximation is legitimate because the lateral
structure of surface is smoothed out by the first solid layer of
helium atoms. The 3-9 potential used in this work is derived from the
silicon-helium interaction \cite{ColeSurfSci91}; we have
supplemented this potential by a 4-10 potential due to averaging
Lennard--Jones 6-12 potentials for the $^4$He-$^4$He interaction
over a plane. Thus, our substrate potential has the form
\begin{equation}
U_{\rm sub}(z) = U_3(z-z0) + U_4(z)\nonumber
\end{equation}
where $U_3(z)$ is the common 3-9 potential \cite{ColeSurfSci91}
\begin{equation}
U_3(z) = \left[{4 C_3^3\over 27 D^2}\right]{1\over z^9} - {C_3\over z^3}
\end{equation}
with a well--depth $D = 128~$K and a range $C_3 = 2000~$K\AA$^{3}$.
The potential $U_4(z)$ has the form
\begin{equation}
U_4(z) = 4\pi\epsilon\rho_1\sigma^2
\left[{1\over5}\left({\sigma/\over z}\right)^{10} - 
{1\over2}\left({\sigma\over z}\right)^4\right]\,.
\end{equation}
Here, $\epsilon = 10.22~$K and $\sigma=2.556~$\AA\ are the usual
deBoer-Michels parameters for the helium-helium interaction. The areal
density of the solid monolayer was taken to be $\rho_1 =
0.07~$\AA$^{\-2}$ and a thickness $z_0 = 3.3~$\AA.  The picture is
consistent values for hectorite gaps \cite{WadaHectoritePRB}, the
areal density is somewhat lower than the one of the {\it second\/}
solid layer on graphite \cite{LauterExeter}.
 
The helium film considered here is thick enough to support layer
rotons; ripplons will also appear for the film model which has a free
surface. In the filled gap case the amount of helium between the Si
planes was chosen to correspond the filling of an aerogel sample
surrounded by bulk liquid in equilibrium. The chemical potential of
the confined liquid and that of a reservoir of bulk liquid turned out
to be equal ($\mu\approx-7.2$~K) at $n=0.85$~\AA$^{-2}$. For
comparison, a 40~\AA\ thick slice of bulk liquid at the equilibrium
density 0.02185~\AA$^{-3}$ would correspond to the coverage
$n=0.87$~\AA$^{-2}$.

Some excitations modes are bulk--like, these have no significant angle
dependence. In the bulk liquid the roton energy decreases and the
roton minimum moves to higher momenta as the density increases. Traces
of this effect are also visible in the 3D roton: the roton energy is
above it's equilibrium bulk value because of the low density regions
between the layers.

Layer modes, on the other hand, are entirely different.  In films,
layer rotons propagate mainly in the highest density liquid layer
closest to the substrate, hence their motion is well confined to two
dimensions and their propagation direction is always parallel to the
surface. Their energy has a parabolic minimum at a fixed wave vector
$\veck_{\|,0}$ in the plane parallel to the surface. Upon changing the
orientation of the surface also the in--plane component of the total
wave vector transfer changes. As a result, the location of the layer
roton minimum shifts.

We will show our results for the dynamic structure function as
gray--scale plots. In these maps of the dynamic structure function,
darker areas correspond to higher value of $S(k,\omega)$. To emphasize
low--intensity regions we use the scale $S(k,\omega)^{1/4}$, the same
gray scale will be used throughout this work. To facilitate plotting
we have also introduced a 0.05~K Lorentzian broadening of the
structure functions. We measure the scattering angle from the
substrate plane, so $\theta = 0$ corresponds to grazing angle and
$\theta = 90$ degrees is scattering perpendicular to the substrate
plane.


\subsection{Thick $^4$He films on silicon}

\begin{figure}[tb]
\vbox{
\includegraphics[width=0.48\textwidth]{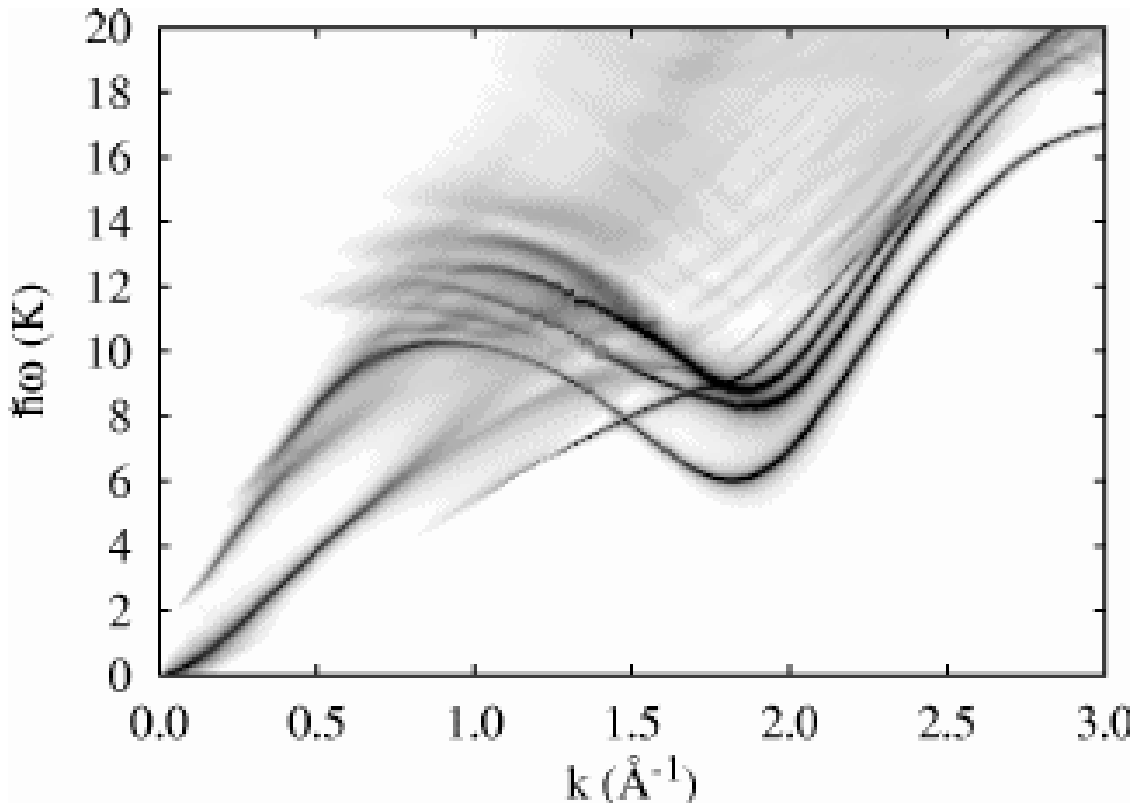}\\
\includegraphics[width=0.48\textwidth]{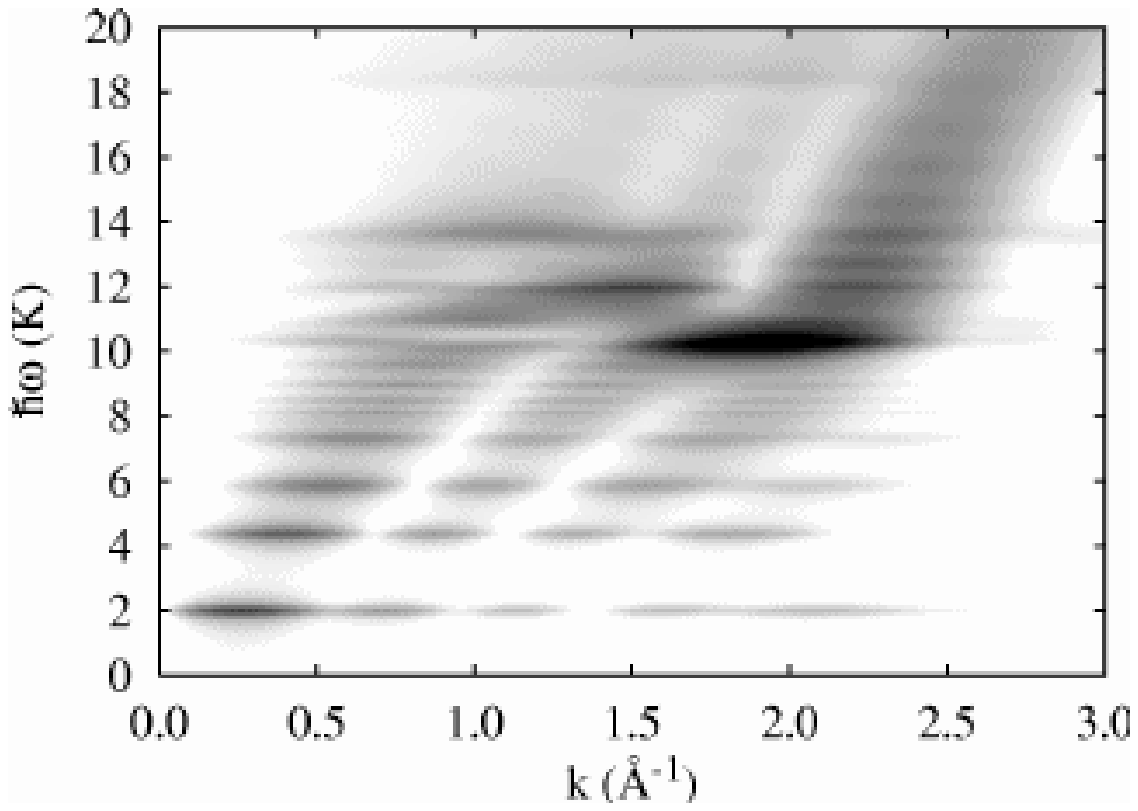}
}
\caption{The dynamic structure function of a thick $^4$He film on a Si
surface (coverage $n = 0.3$~\AA$^{-2}$) at grazing angle (upper panel)
and perpendicular to the substrate (lower panel).}
\label{fig:skw-film-si300-00+90}
\end{figure}

Fig.~\ref{fig:skw-film-si300-00+90} shows the dynamic structure
function of a thick $^4$He film on Si for scattering angles
$\theta=$~0 and $\theta = 90$~degrees. The former shows a ripplon mode
with dispersion $\hbar\omega(k)\sim k^{3/2}$, and also two layer
rotons below the bulk roton are clearly distinguishable. Perpendicular
scattering shows a discrete set of low--energy dispersionless modes,
but some strong resonances appear also above the continuum limit $-\mu
= 7.7$~K. In the perpendicular case the familiar phonon--roton form is
still recognizable, but beside it there are secondary maxima, which
are roughly evenly spaced in ${\bf k}_\perp$. These strands appear
because the projection onto plane waves in Eq.(\ref{eq:Sqwfilm}) gives
contributions to multiple perpendicular momenta ${\bf k}_\perp$, and
the dynamic structure function shows features coming from the Fourier
transform of the density profile. A detailed description of how the
strands appear was recently given in
Ref.~\onlinecite{apaja-godfrin-krotscheck-lauter-01}.

\begin{figure}[tb]
\includegraphics[width=0.48\textwidth]{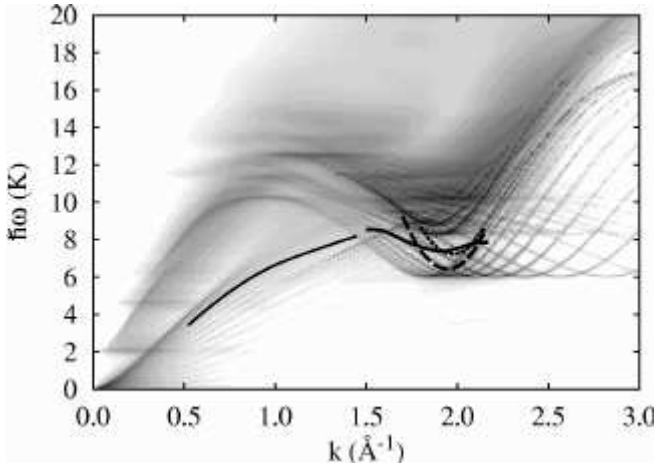}
\caption{The dynamic structure function of a thick film on a Si
surface (coverage $n = 0.3$~\AA$^{-2}$). The plot shows the sum of
contributions at angles between 0 and 90 degrees with 5 degree
steps. The dashed lines show the experimental layer roton dispersion
for aerogel (upper curve) and Vycor (lower curve) by Plantevin {\em et
al.}~\cite{plantevin-etal-01}; The solid curve shows the layer roton and
ripplon data from Ref. \onlinecite{LauterAerogel}.}
\label{fig:skw-film-si300-sum}
\end{figure}

Randomly oriented surfaces give rise to an angular averaged response.
Fig.~\ref{fig:skw-film-si300-sum} shows such a theoretical dynamic
structure function, where contributions from several scattering angles
between 0 and 90 degrees have been added. While the contributions from
bulk--like rotons add up to a single curve, the ones from 2D modes do
not. The low--$k$ region shows faints steps at about
$\hbar\omega=$~0~K, 1.9~K, 4.4~K, 6.0~K etc., reminiscent of the
states turning dispersionless as the angle increases; hence the
fan--like structures. The strongest of these comes from the ripplons,
and is probably the only one that can be in experiments.  As mentioned
earlier, the layer roton minimum shifts to higher total momentum
transfers as the scattering angle increases. This has not been seen in
neutron scattering experiments, which show a clear parabolic
dispersion for the layer modes. The parallel direction gives the the
strongest signal in neutron scattering, and we can only assume that
the contribution from other angles in the raw data is simply too small
to give any information to be inverted to $S(k,\omega)$.

\begin{figure}[tb]
\includegraphics[width=0.48\textwidth]{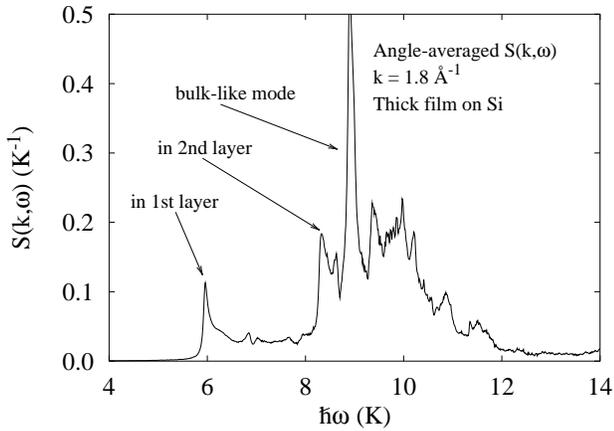}
\caption{The fully angle--averaged $S(k,\omega)$ of a thick film at
$k=1.8$~\AA$^{-1}$.  The coverage is $0.3$~\AA$^{-2}$.}
\label{fig:skw-film-si300-sum-k1.8}
\end{figure}

\begin{figure}[p]
\vbox{
\includegraphics[width=0.46\textwidth]{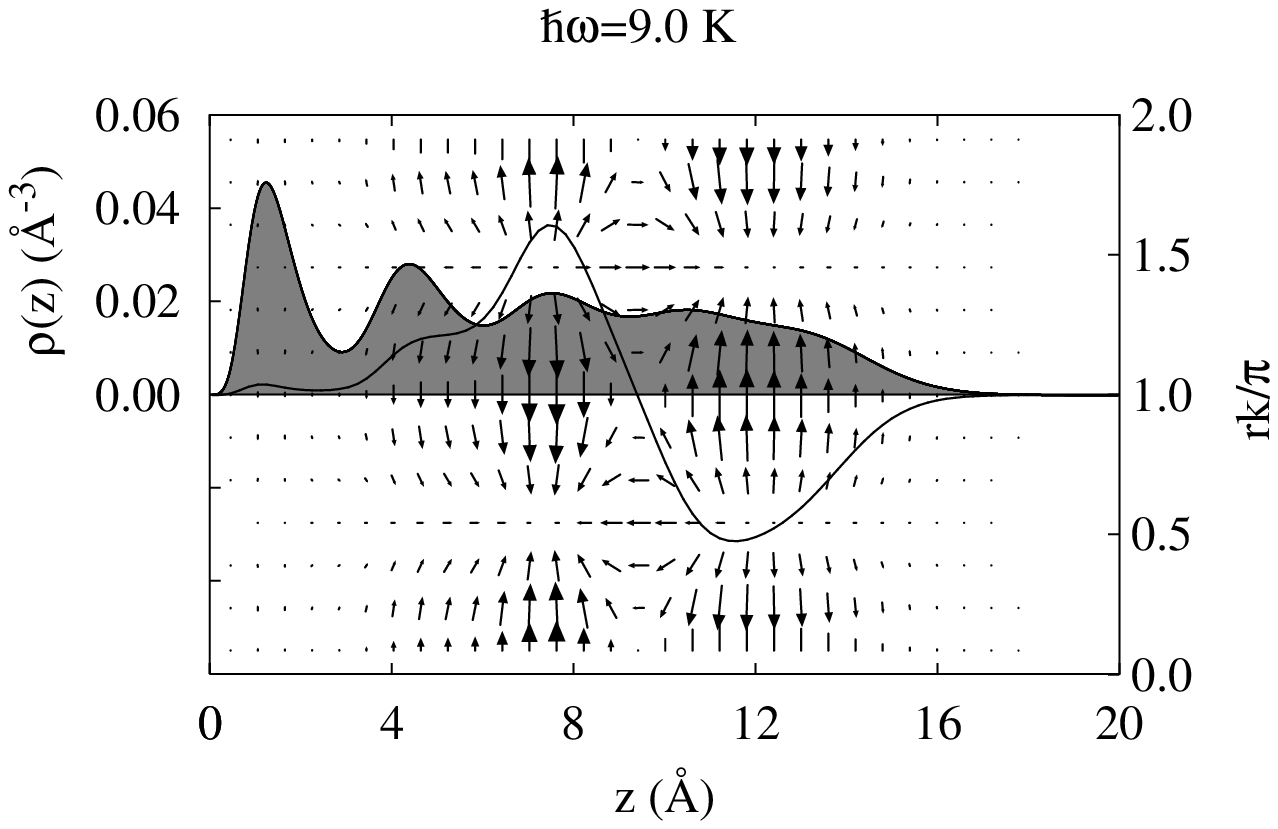} \\
\includegraphics[width=0.46\textwidth]{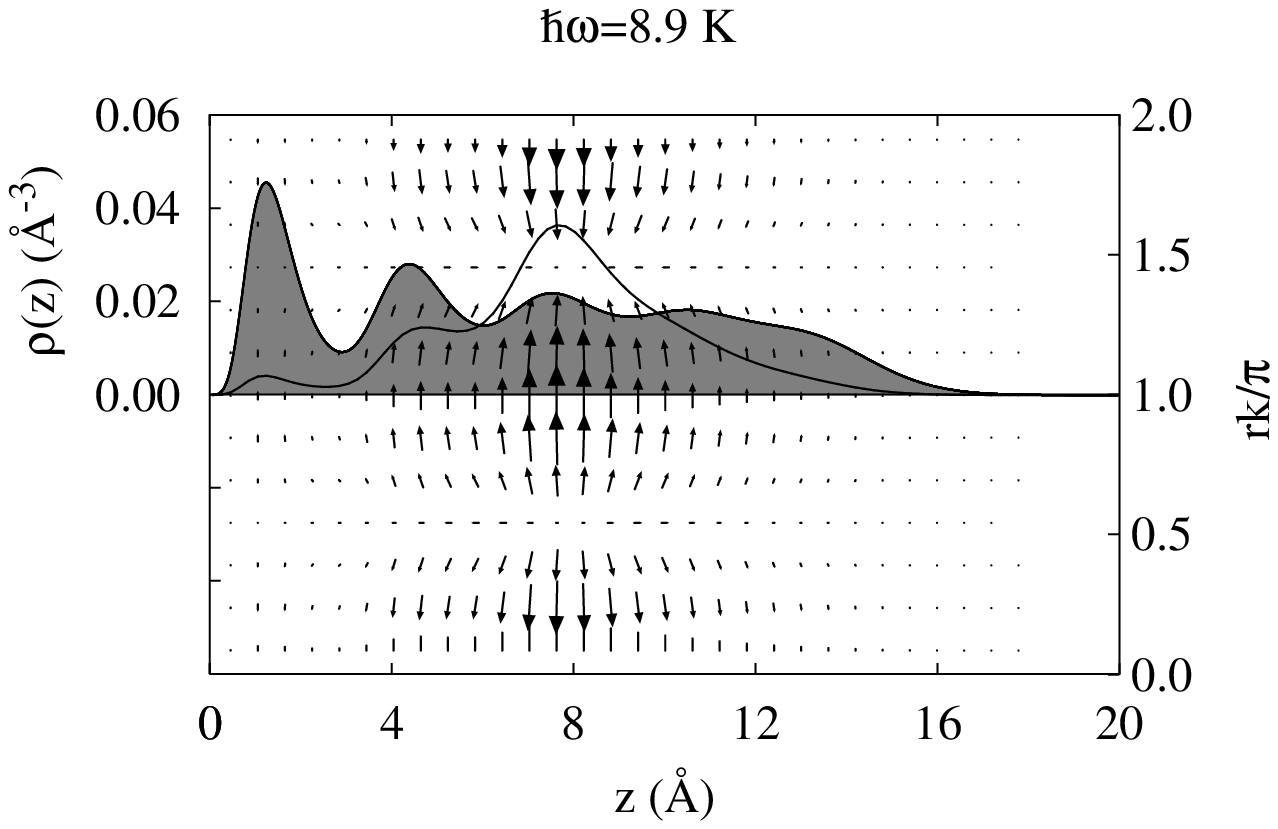} \\
\includegraphics[width=0.46\textwidth]{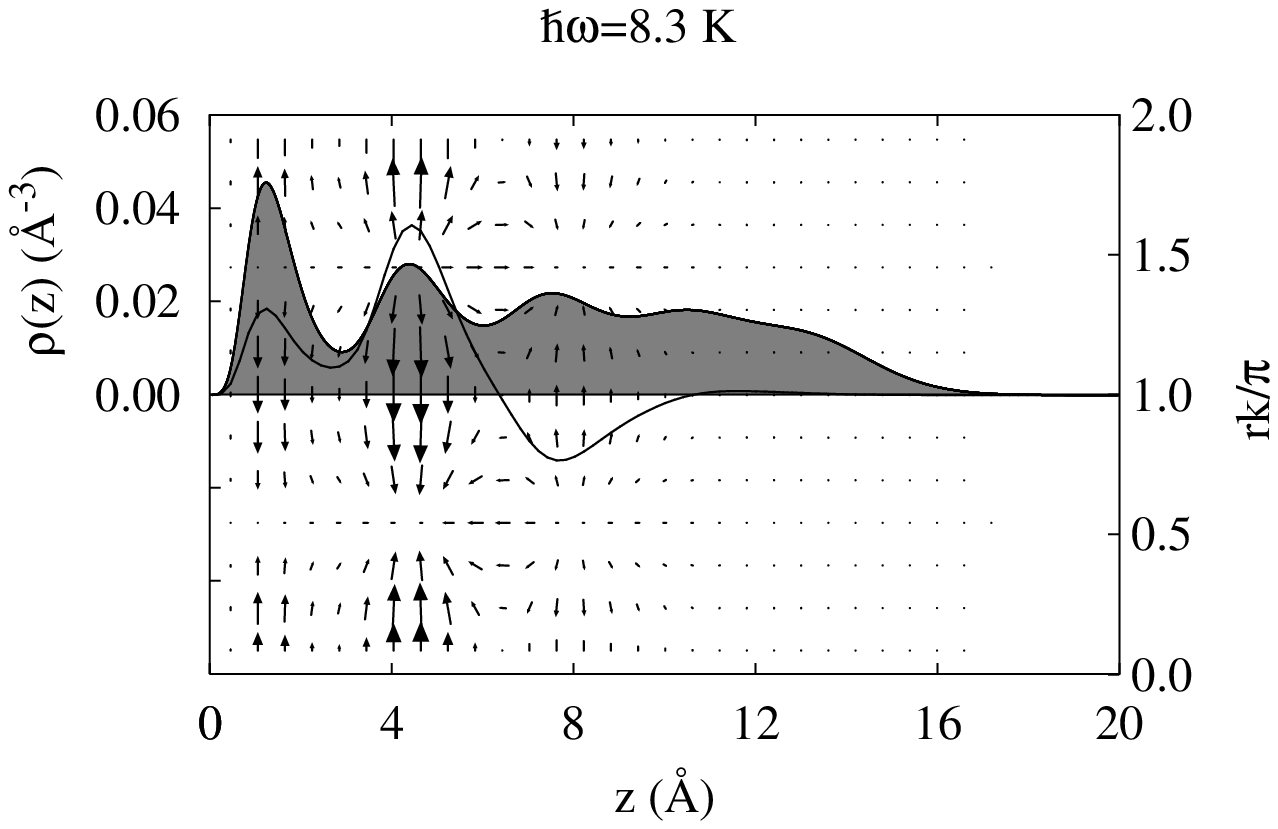} \\
\includegraphics[width=0.46\textwidth]{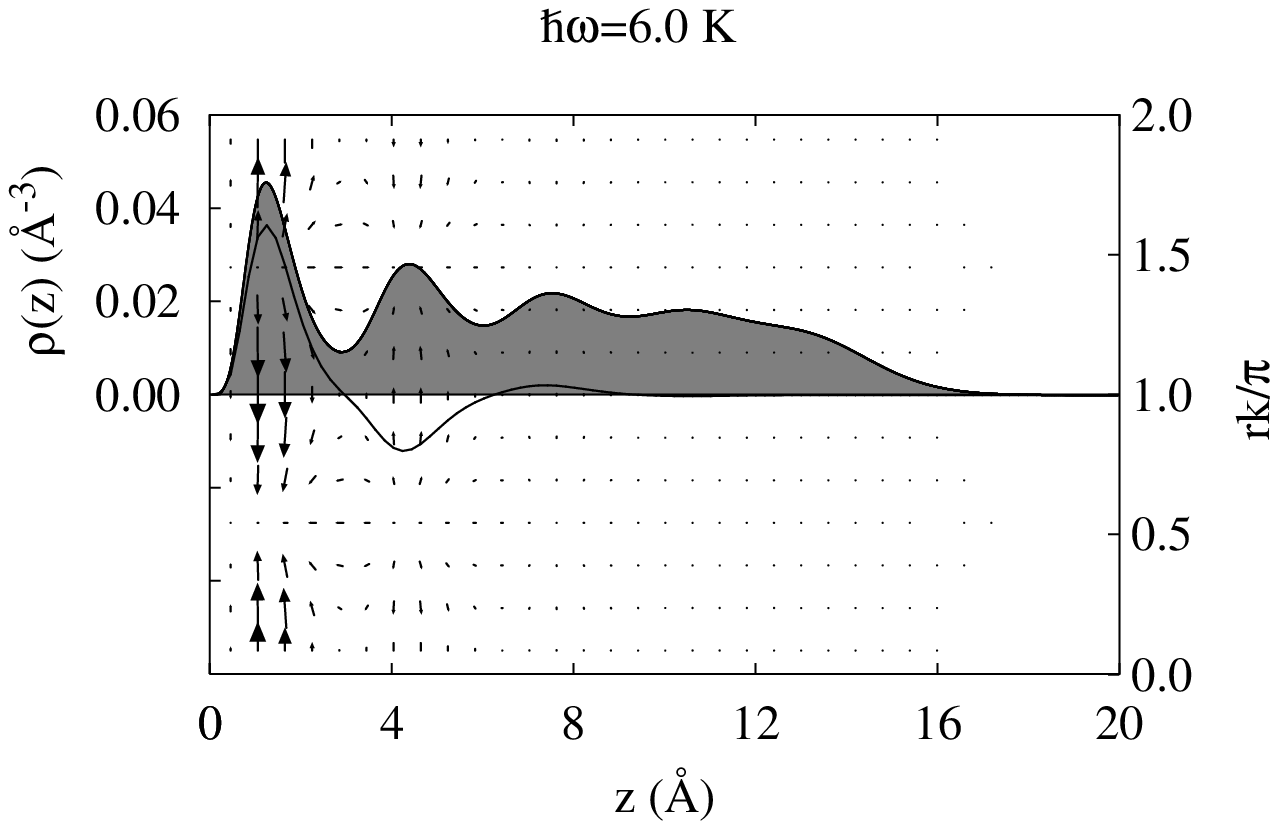} 
}
\caption{Transition density (thick line) and transition current
(arrows; right scale) for the low--energy modes at $k=1.8$~\AA$^{-1}$
for a thick film on Si substrate and scattering parallel to the
surfaces, $\theta = 0$. The grayscale plot shows the density
profile. Excitations energies are indicated in the figures, the lowest
panel corresponds to lowest energy.}
\label{fig:trans-film-si300-0}
\end{figure}

Fig.~\ref{fig:skw-film-si300-sum-k1.8} shows a fully angle--averaged
dynamic structure function at $k=1.8$~\AA$^{-1}$, near the roton
minimum. Angle averaging causes the layer mode peaks to become
asymmetric, while it has no effect on the bulk--like modes. The
asymmetry is clearly visible in the first--layer peak, which has a
broad tail on the high--energy side. This broadening is just due to
summing up of the multiple layer roton parabola, which show as separate
curves in Fig.~\ref{fig:skw-film-si300-sum}.
A direct identification of the modes is given by the transition
density and current, shown for the thick film in
Fig.~\ref{fig:trans-film-si300-0} for $\theta = 0$. The scale on the
right corresponds to the transition current, depicted as a vector
field, for the duration of one oscillatory cycle which takes about
$48\times\hbar\omega$K$^{-1}$ picoseconds. The lowest energy mode for
grazing angle is a layer roton, which involves atoms in the layer
closest to the substrate. The next mode is loosely confined to the
second layer, but subsequent modes spread throughout the liquid as
there are no more well--separated layers available. From the third
mode on the excitations are three dimensional, as well as they can be
in a 15~\AA\ thick film.

\begin{figure}[p]
\vbox{
\includegraphics[width=0.46\textwidth]{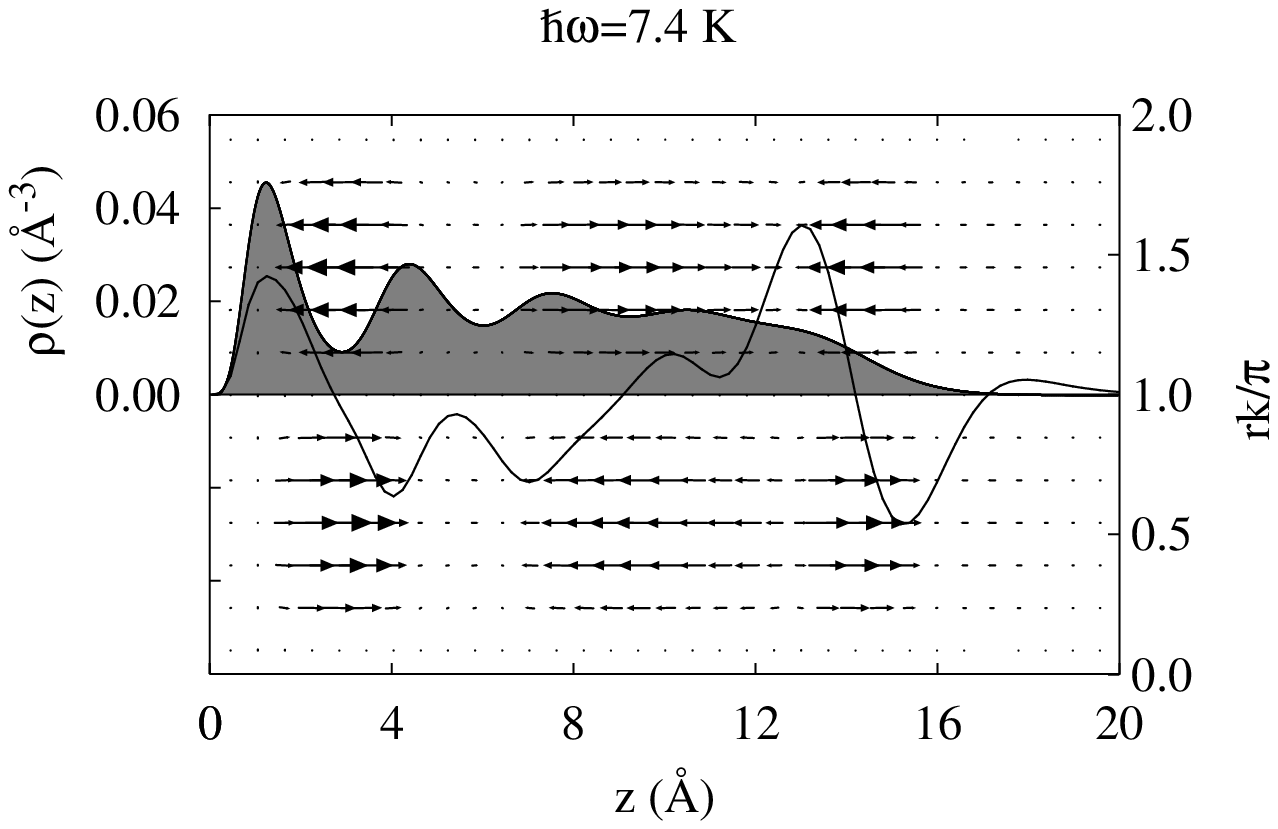} \\
\includegraphics[width=0.46\textwidth]{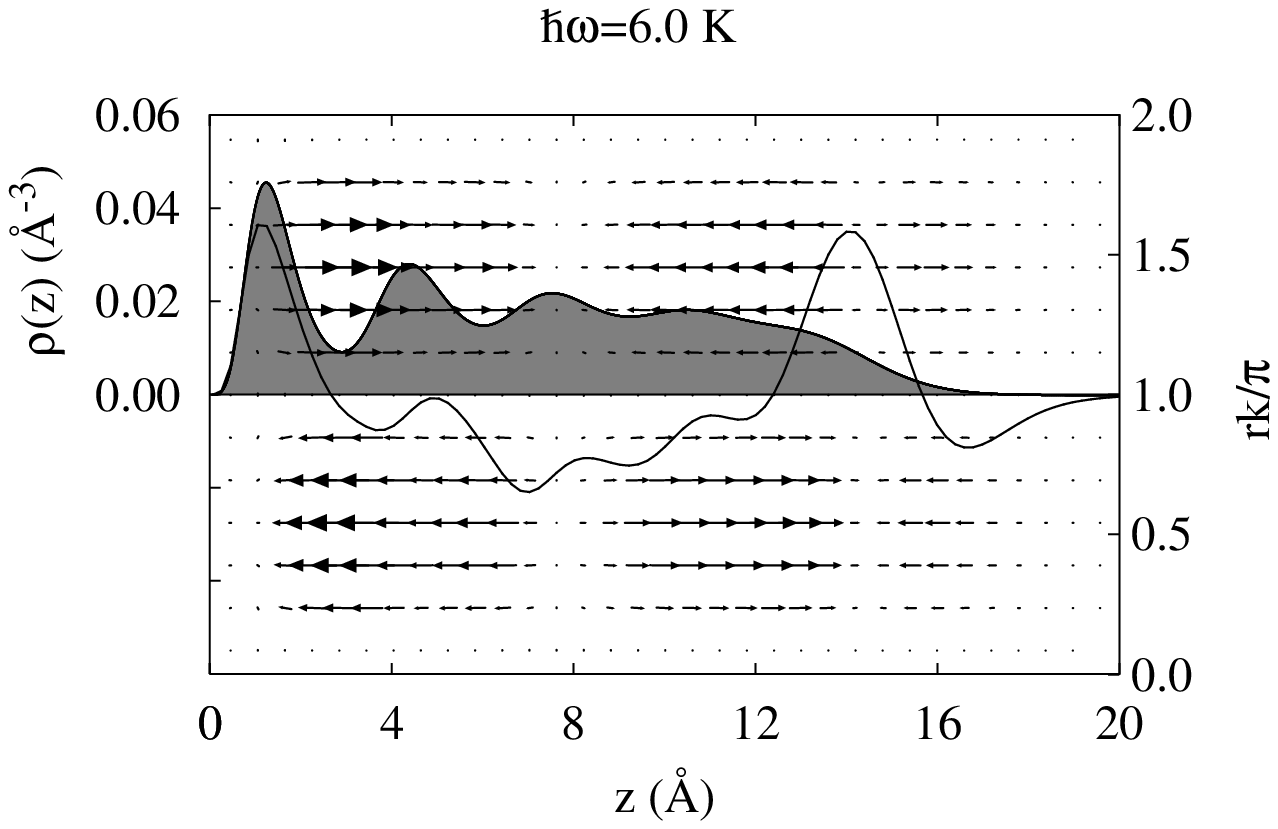} \\
\includegraphics[width=0.46\textwidth]{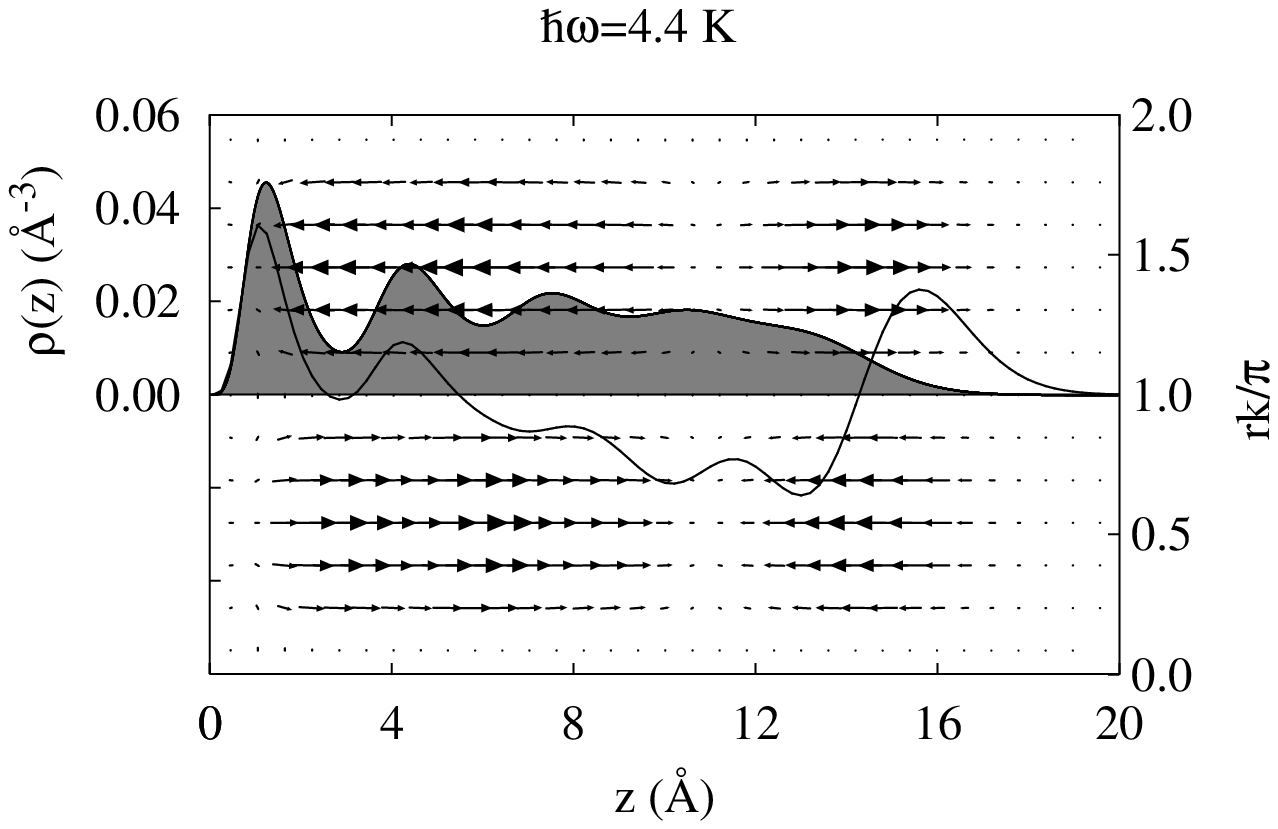} \\
\includegraphics[width=0.46\textwidth]{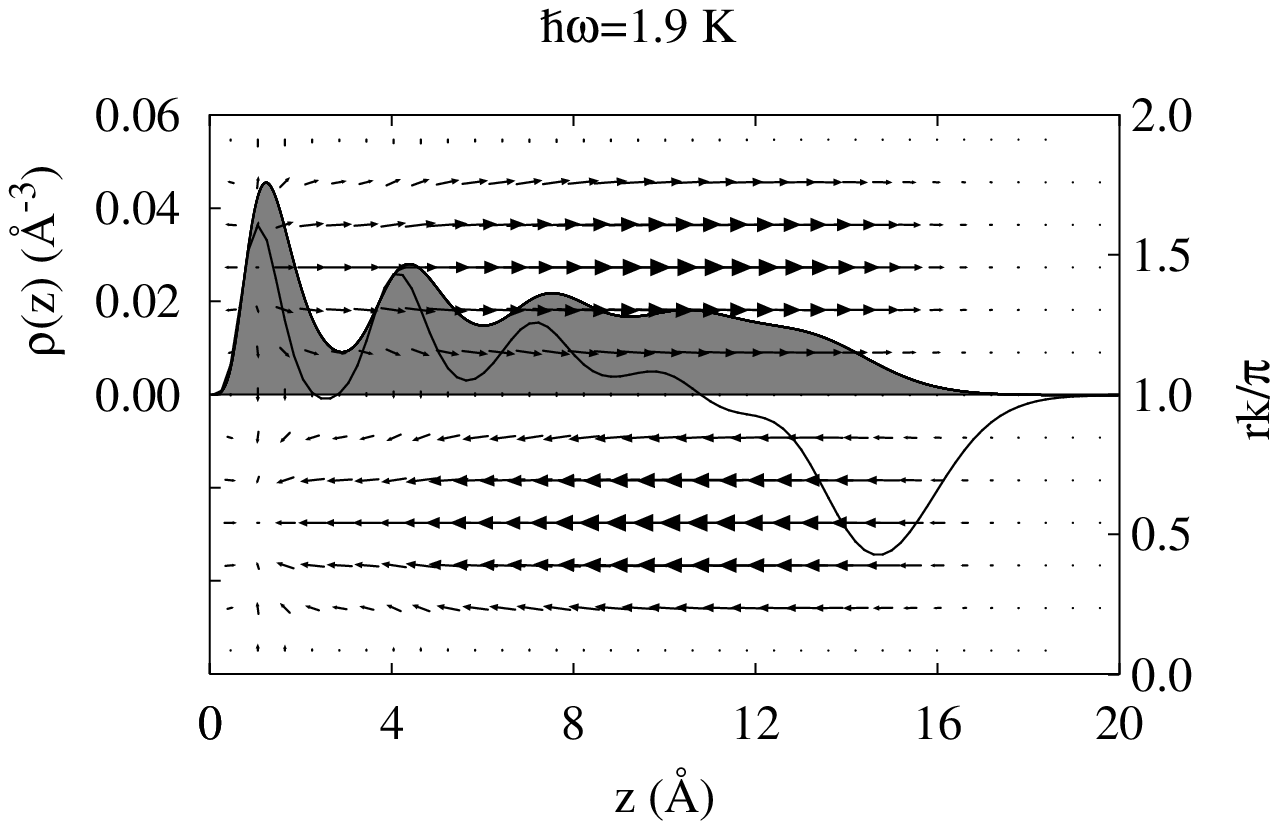} 
}
\caption{Same as Fig.~\protect\ref{fig:trans-film-si300-0} for $\theta
= 90$ (specular scattering).}
\label{fig:trans-film-si300-90}
\end{figure}

The situation in the perpendicular scattering is completely
different. The lowest panel of Fig.~\ref{fig:trans-film-si300-90}
shows the mode at $\hbar\omega =1.9$~K, which already has one node in
the transition density. Particle number conservation forbids a
nodeless mode, so the first real excitation is one where the liquid
oscillates with respect to the substrate.

Fig.~\ref{fig:trans-film-si300-90} can be looked upon as a pictorial
view of quantum evaporation. Higher energy modes involve motion of
atoms near the surface, and finally modes with energy above minus the
chemical potential (now $\mu = -7.7$~K) are energetic enough to kick
out atoms to continuum states.

\subsection{$^4$He filling space between two silicon surfaces}

\begin{figure}[tb]
\includegraphics[width=0.48\textwidth]{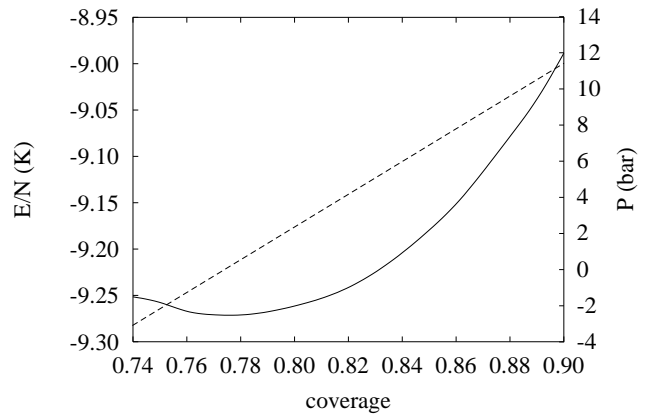}
\caption{The total energy (solid line, left scale) and pressure
(dashed line, right scale) of a helium filled 40~\AA\ wide gap as a
function of coverage. Liquid in the middle of the gap becomes unstable
at about $0.68$~\AA$^{-2}$ indicating the formation of two separate
thick films on both walls.}
\label{fig:ener-pres}
\end{figure}

Fig.~\ref{fig:ener-pres} shows the energy and pressure in the case,
where helium fills the space between the substrate walls. The
equilibrium density of the filled gap by itself would be at about
$\rho=0.78$~\AA$^{-2}$, but the liquid is in balance with an external
bulk $^4$He reservoir at $\rho=0.85$~\AA$^{-2}$.  Below
0.68~\AA$^{-2}$ the liquid is no longer translationally invariant
parallel to the walls, and the instability in the liquid solution is
an indication of capillary condensation. At low density liquid forms
two films covering both surfaces. In the filled gap case, if the
density of helium between the walls is further increased, one observes
a sequence of layering transitions: the number of helium layers
increases. This filling scenario has been discussed in detail in
Ref.~\cite{apaja-krotscheck-01b}.

\begin{figure}
\includegraphics[width=0.48\textwidth]{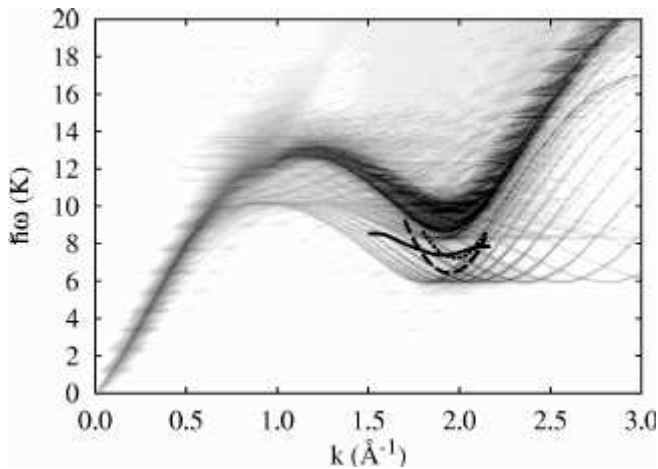}
\caption{The dynamic structure function of the filled gap.  The plot
shows the sum of contributions at angles between 0 and 90 degrees in 5
degree steps. The calculated layer roton shifts to higher $k$ as a
function of angle. The dashed lines show the experimental layer roton
dispersion for aerogel (upper curve) and Vycor (lower curve) by
Plantevin {\em et al.}~\cite{plantevin-etal-01}; the solid curve 
shows results from Ref. \onlinecite{LauterAerogel}.}
\label{fig:skw-gap}
\end{figure}

The (partially) angle averaged $S(k,\omega)$ in Fig.~\ref{fig:skw-gap}
is qualitatively different from the film case result in
Fig.~\ref{fig:skw-film-si300-sum} only in the low--momentum
region. There is now only one strong maximum corresponding to bulk
phonons on top of a faint step structure. The steps are a remainder of
the discrete states in the perpendicular or nearly perpendicular
directions. The angle averaged $S(k,\omega)$ near the roton wave
vector is quite similar to the one in the thick film case, depicted in
Fig.~\ref{fig:skw-film-si300-sum-k1.8}

\begin{figure}
\vbox{
\includegraphics[width=0.46\textwidth]{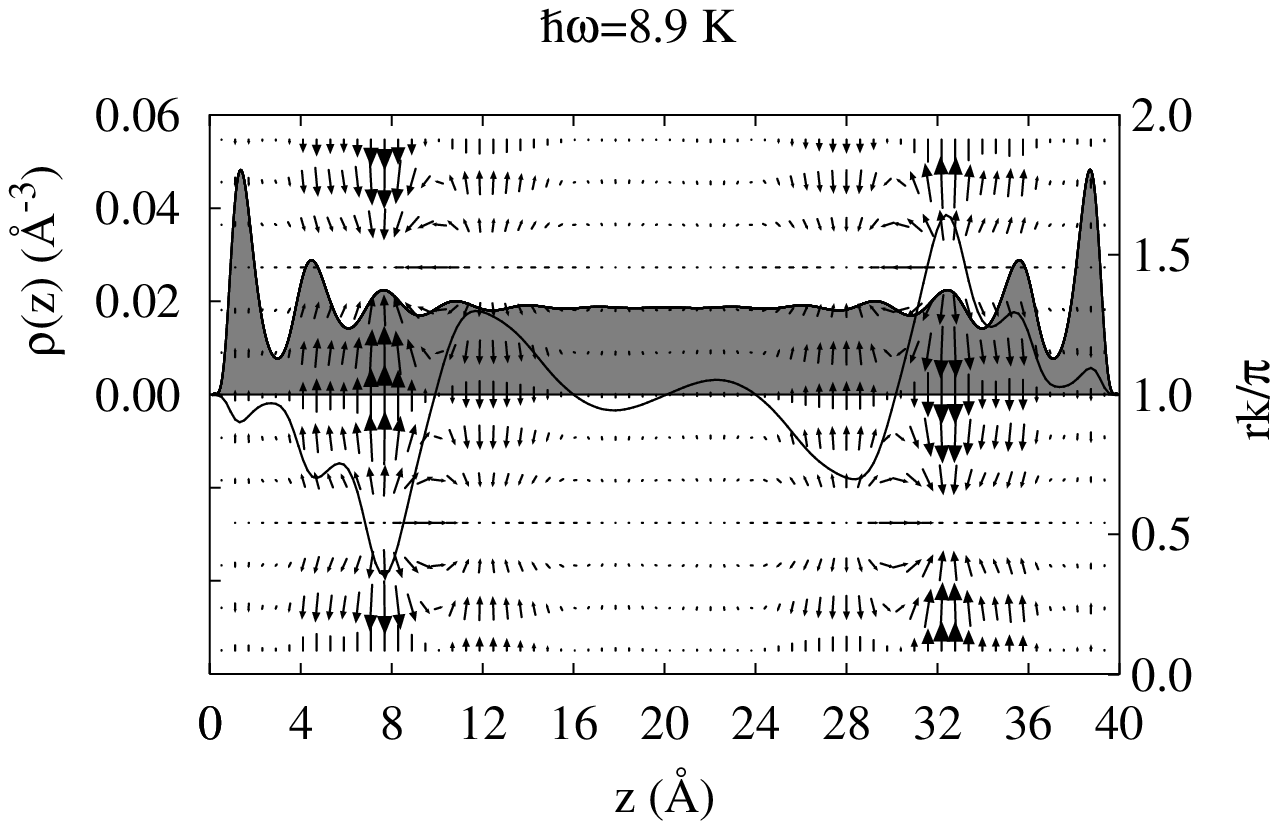} \\
\includegraphics[width=0.46\textwidth]{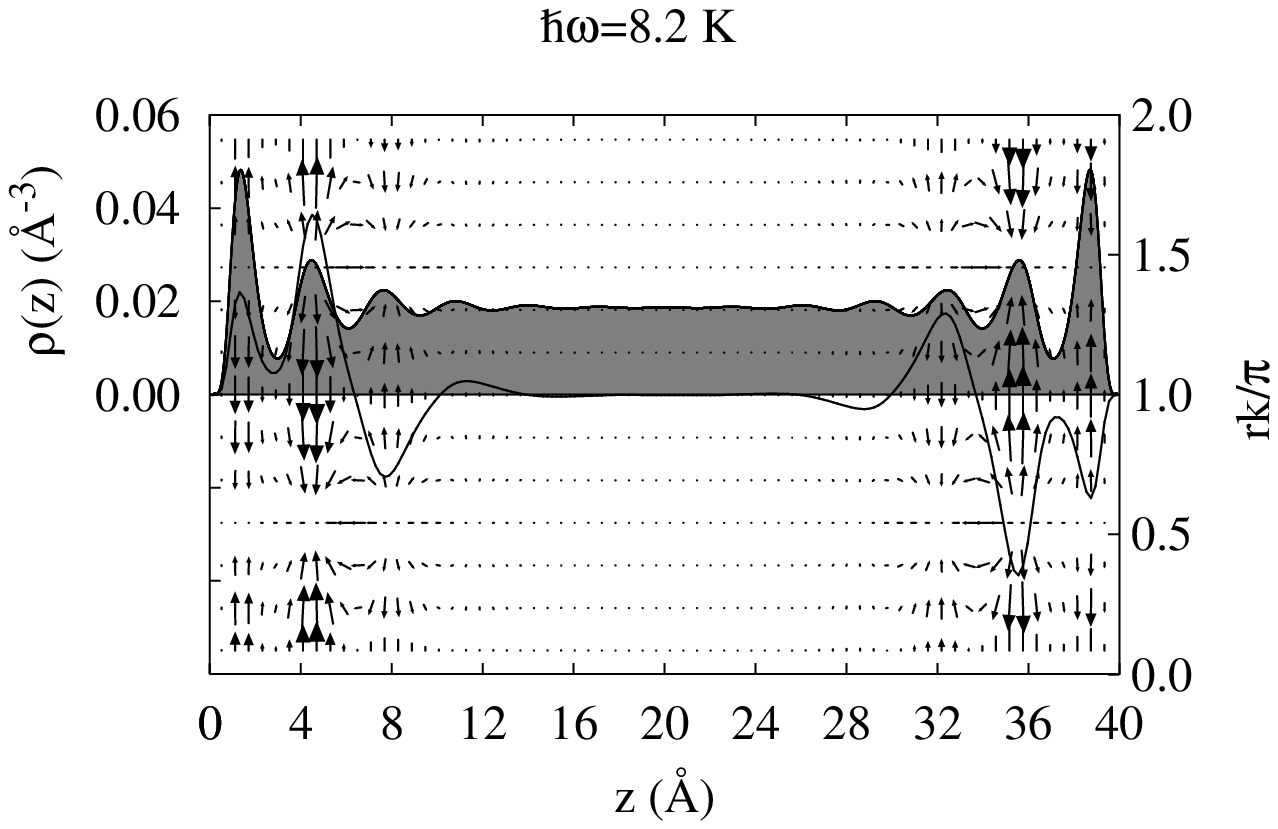} \\
\includegraphics[width=0.46\textwidth]{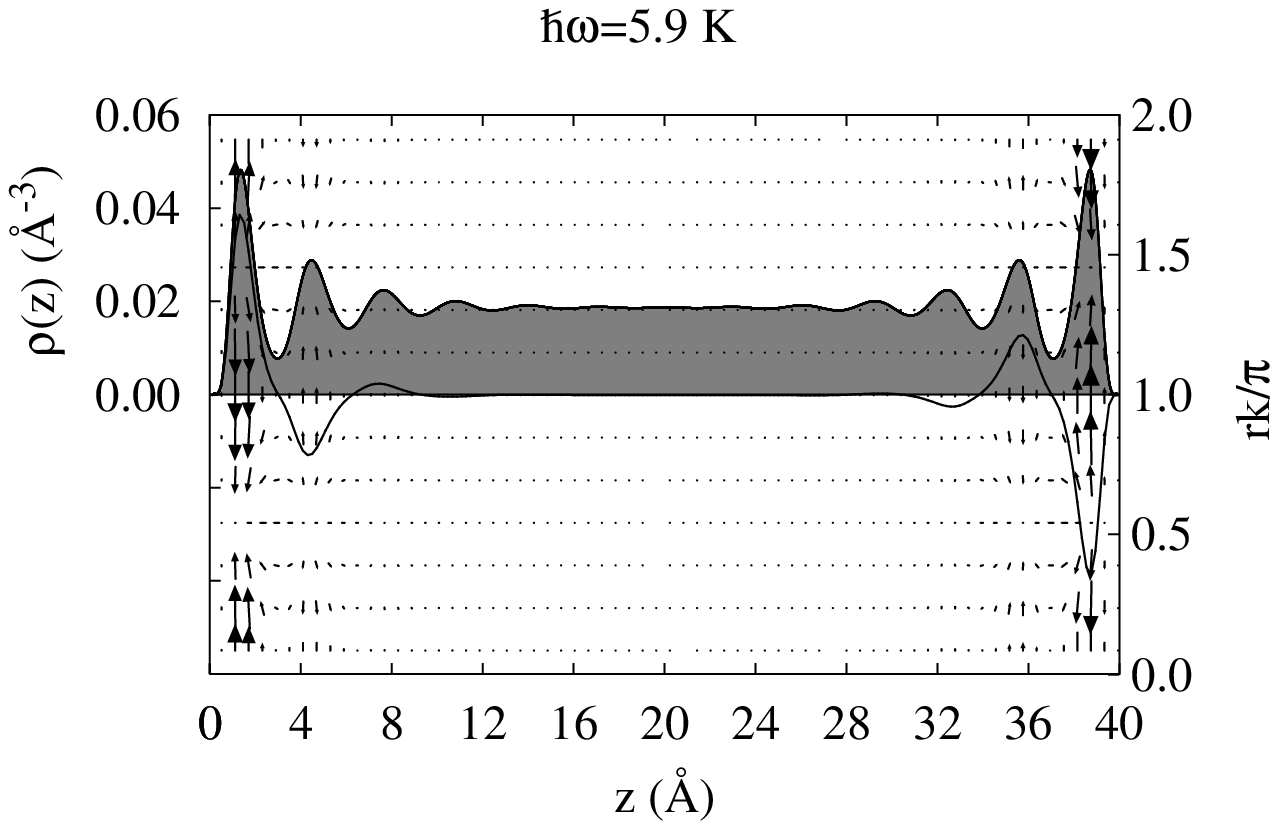} 
}
\caption{Transition density (thick line) and transition current
(arrows; right scale) in the low--energy modes at $k=1.8$~\AA$^{-1}$
in the helium filled gap at $n=0.85$~\AA$^{-2}$ for grazing angle. The
grayscale plot shows the density profile. The transition density has
arbitrary scale.  Excitations energies are indicated in the figures,
lowest panel corresponds to lowest energy. Due to the symmetric
density profile there is pairwise degeneracy of modes corresponding to
in--phase and out--of--phase oscillation between the left and the
right side; we show here only the latter ones.}
\label{fig:trans-squeezed-0}
\end{figure}

\begin{figure}
\vbox{
\includegraphics[width=0.46\textwidth]{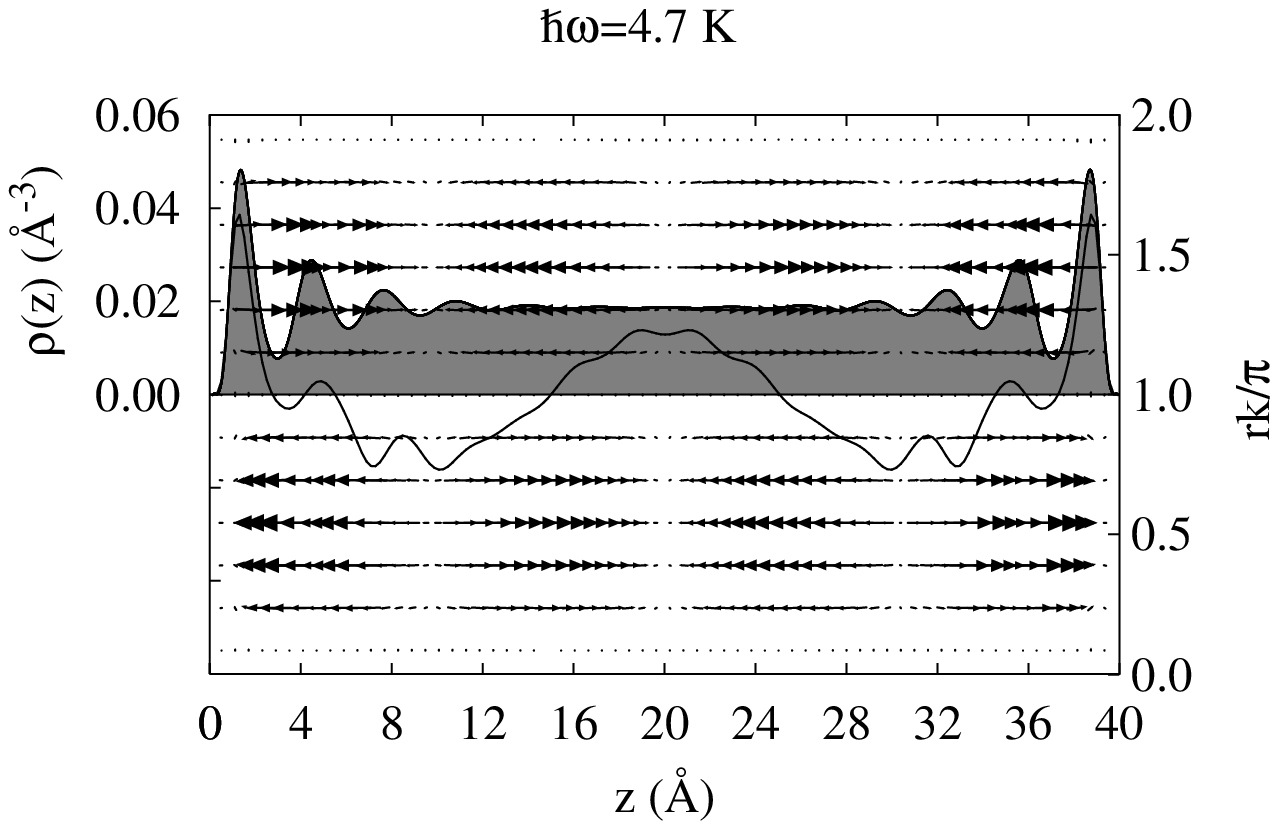} \\
\includegraphics[width=0.46\textwidth]{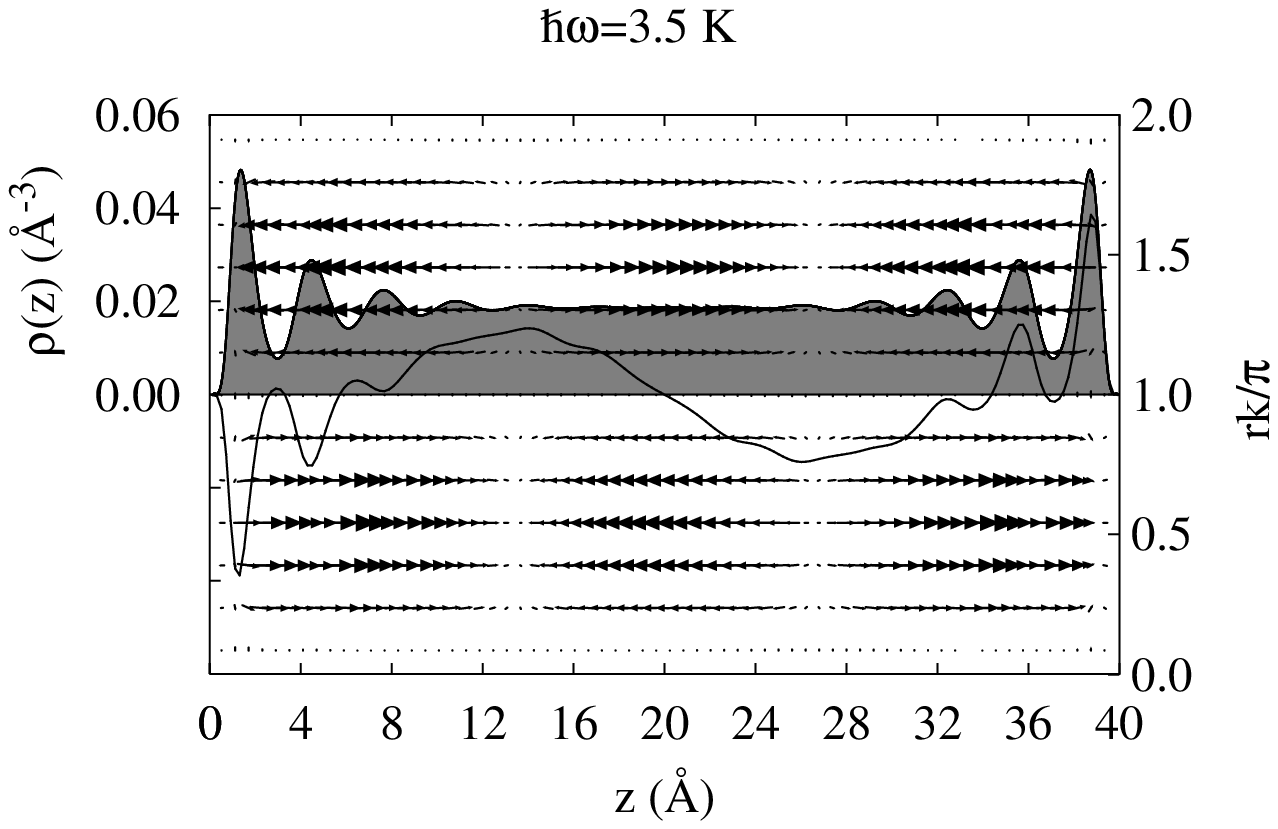} \\
\includegraphics[width=0.46\textwidth]{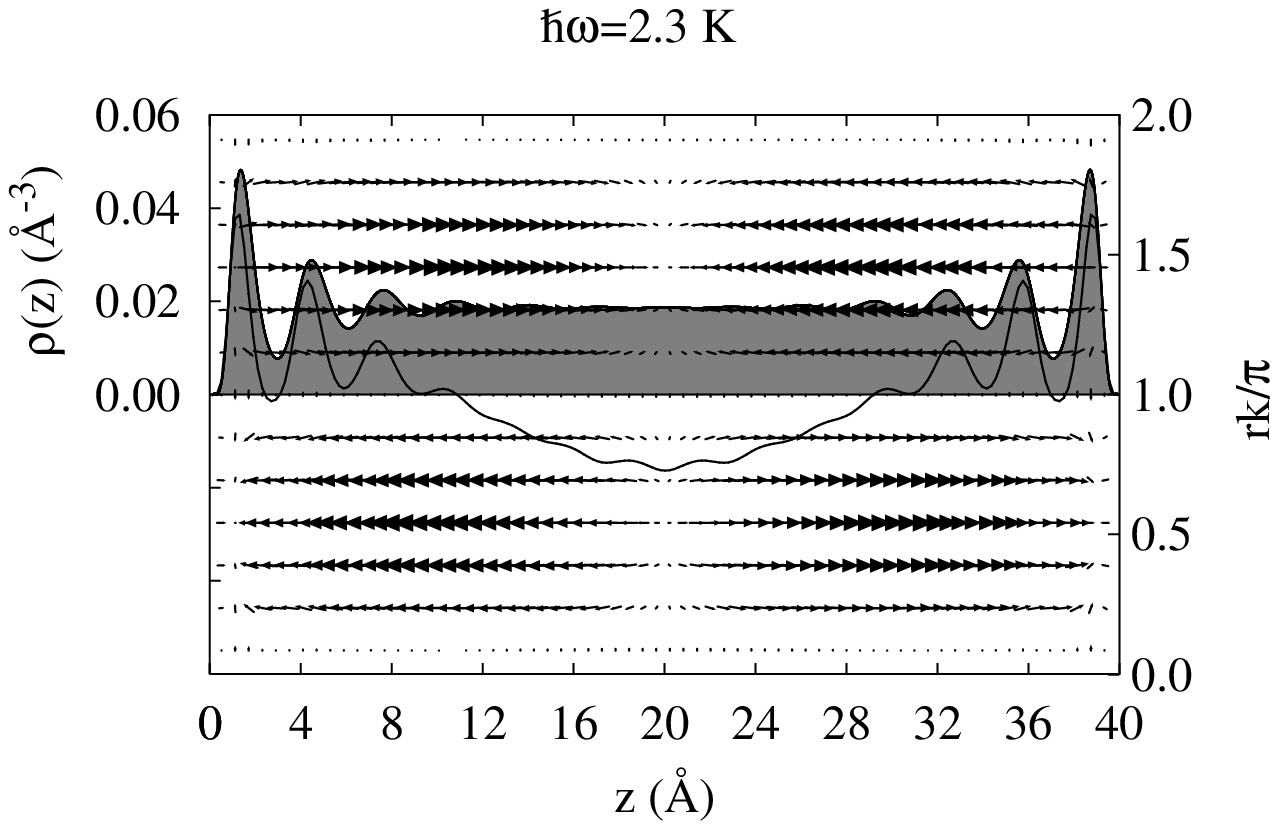} \\
\includegraphics[width=0.46\textwidth]{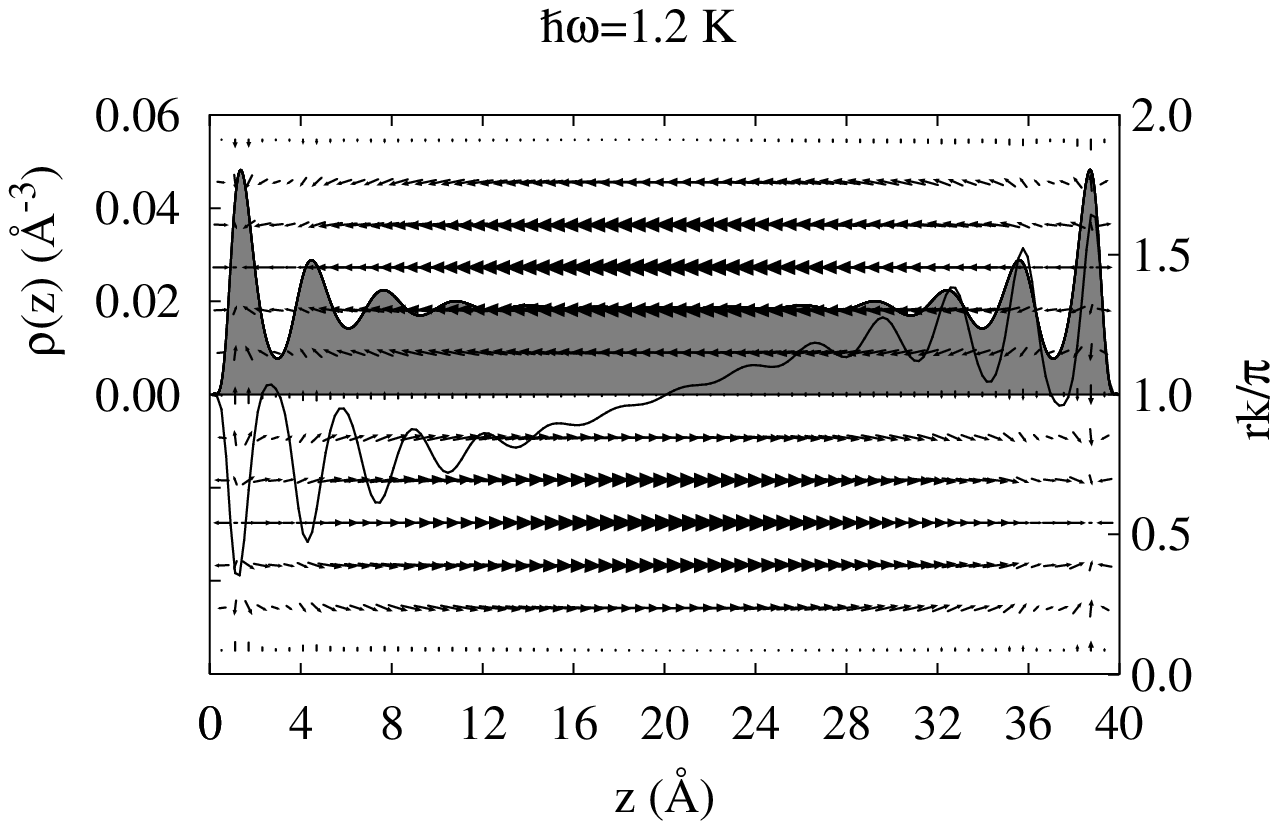} 
}
\caption{Same as Fig.~\protect\ref{fig:trans-squeezed-0} for the 90 degree
angle (specular scattering), except that now there is no degeneracy
and we show all low--energy modes.}
\label{fig:trans-squeezed-90}
\end{figure}

The transition densities and currents are shown in
Figs.~\ref{fig:trans-squeezed-0} and \ref{fig:trans-squeezed-90} for
scattering angles 0 and 90 degrees, respectively. At grazing angle the
low--energy excitation reside in the layers near the substrate, and
pairs of modes corresponding to in-- and out--of--phase oscillations
are degenerate. The modes with even number of oscillations have their
counterparts in the case of two films on both walls, as one can see by
comparing Figs.~\ref{fig:trans-film-si300-90} and
\ref{fig:trans-squeezed-90}. The lowest mode depicted in
Fig.~\ref{fig:trans-squeezed-90} with odd number of oscillations have
liquid oscillating back and forth in the middle of the gap, and with
decreasing density this oscillation is finally able to divide the
system into two films. The point where this mode becomes soft is the
spinodal instability of the filled case.

\section{Conclusions}

In order to model \he4 confined in materials like aerogel, Vycor or
Geltech, we have calculated the energetics, the density profile and
the dynamical properties of thick \he4 film on Si substrate and
between two substrate planes. The simulation of confinement in
spherical pores, which is perhaps the model that is closest to
reality, is also feasible with our techniques, but computationally
less efficient. For the excitations near the roton wave vector, where
the most interesting phenomena are occurring, the use of a simplified
planar geometry should not make much difference because the roton wave
length is much shorter than any lengthscale on which the lateral
structure of the aerogel changes. The two--wall geometry has actually
been used in specific heat measurements by Gasparini and collaborators
\cite{rhee-gasparini-bishop-89,
mehta-kimball-gasparini-98,mehta-kimball-gasparini-99,kimball-gasparini-00},
but the distance between the silicon planes was a couple of magnitudes
larger than in our model. For a large wall distance the mid section of
the liquid is essentially unaware of the substrate. Thus one would
have a thick region of basic bulk helium, which would show up almost
exclusively in the dynamic structure function.

The results for thick helium films are relevant for large pores at low
filling, when opposite walls are approximately noninteracting and the
curvature of the liquid can be ignored. We have not considered the
possibility of unevenly filled pores or a distribution pore sizes,
because the characteristics of porous systems are very material
dependent, and our aim was to discuss phenomena that are universal
in such materials.

There are presently two sets of data for the layer roton, those of
Refs. \onlinecite{plantevin-etal-01} and of
\onlinecite{LauterAerogel}. When comparing theory and experiment, two
aspects are of interest: One is the energy of the layer roton, and the
other one the dispersion curve. As can be seen from the measurements of
Ref. \onlinecite{plantevin-etal-01} and \onlinecite{LauterAerogel}
shown in Figs.  \ref{fig:skw-film-si300-sum} and \ref{fig:skw-gap}, as
well as from other measurements
\cite{fak-plantevin-glyde-mulders-00,glyde-etal-00}, that the energy
of the layer roton depends sensitively on the nature of the
substrate. This is expected because different substrates will lead to
different densities of the first liquid layers and, hence, to
different phonon--roton spectra. Indeed, the dependence is so
sensitive that it might be possible to determine the strength of the
substrate potential from the energy of the roton gap.

The second aspect is the shape of the dispersion curve. The
measurements of Ref. \onlinecite{LauterAerogel} show a significantly
smaller curvature of the dispersion curve than those of the other
experiments, but they agree better with our calculations. Keeping the
extreme difficulty of extracting these data from neutron scattering
spectra in mind, we tend towards the view that the experiments of Ref.
\onlinecite{LauterAerogel} provide a more accurate description.  This
is so for two reasons: First, the curvature of the roton in purely
two--dimensional $^4$He is smaller than that obtained in Ref.
\onlinecite{plantevin-etal-01}, and second, as shown in Figs.
\ref{fig:skw-film-si300-sum} and \ref{fig:skw-gap}, the angular
averaging leads to a further broadening.

\begin{acknowledgments}
This work was supported by the Austrian Science Fund (FWF) under
project P12832-TPH. One of us (VA) would like to thank the Department
of Physics at the Technical University of Denmark, where part of this
work was done, for hospitality.
\end{acknowledgments}

\bibliography{papers,papers_vesa}

\end{document}